# Higher Order Programming to Mine Knowledge for a Modern Medical Expert System

**Nittaya Kerdprasop and Kittisak Kerdprasop**

Data Engineering and Knowledge Discovery (DEKD) Research Unit,
School of Computer Engineering, Suranaree University of Technology,
Nakhon Ratchasima 30000, Thailand

**Abstract**

Knowledge mining is the process of deriving new and useful knowledge from vast volumes of data and background knowledge. Modern healthcare organizations regularly generate huge amount of electronic data stored in the databases. These data are a valuable resource for mining useful knowledge to help medical practitioners making appropriate and accurate decision on the diagnosis and treatment of diseases. In this paper, we propose the design of a novel medical expert system based on a logic-programming framework. The proposed system includes a knowledge-mining component as a repertoire of tools for discovering useful knowledge. The implementation of classification and association mining tools based on the higher order and meta-level programming schemes using Prolog has been presented to express the power of logic-based language. Such language also provides a pattern matching facility, which is an essential function for the development of knowledge-intensive tasks. Besides the major goal of medical decision support, the knowledge discovered by our logic-based knowledge-mining component can also be deployed as background knowledge to pre-treatment data from other sources as well as to guard the data repositories against constraint violation. A framework for knowledge deployment is also presented.

## 1. Introduction

Knowledge is a valuable asset to most organizations as a substantial source to support better decisions and thus to enhance organizational competency. Researchers and practitioners in the area of knowledge management view knowledge in a broad sense as a state of mind, an object, a process, an access to information, or a capability [2, 13]. The term *knowledge asset* [24, 26] is used to refer to any organizational intangible property related to knowledge such as know-how, expertise, intellectual property. In clinical companies and computerized healthcare organizations knowledge assets include order sets, drug-drug interaction rules, guidelines for practitioners, and clinical protocols [12].

Knowledge assets can be stored in data repositories either in implicit or explicit form. Explicit knowledge can be managed through the existing tools available in the current database technology. Implicit knowledge, on the contrary, is harder to achieve and retrieve. Specific tools and suitable environments are needed to extract such knowledge.

Implicit knowledge acquisition can be achieved through the availability of the knowledge-mining system. *Knowledge mining* is the discovery of hidden knowledge stored possibly in various forms and places in large data repositories. In health and medical domains, knowledge has been discovered in different forms such as association rules, classification trees, clustering means, trend or temporal patterns [27]. The discovered knowledge facilitates expert decision support, diagnosis and prediction. It is the current trend in the design and development of decision support systems [3, 16, 20, 31] to incorporate knowledge discovery as a tool to extract implicit information.

In this paper we present the design of a medical expert system and the implementation of knowledge mining component. Medical data mining is an emerging area of computational intelligence applied to automatically analyze electronic medical records and health databases. The non-hypothesis driven analysis approach of data mining technology can induce knowledge from clinical data repositories and health databases. Induced knowledge such as breast cancer recurrence conditions or diabetes implication is important not only to increase accurate diagnosis and successful treatment, but also to enhance safety and reduce medication-related errors.

A rapid prototyping of the proposed system is demonstrated in the paper to highlight the fact that higher order and meta-level programming are suitable schemes to implement a complex knowledge-intensive system. For such a complicated system program coding should be done declaratively at a high abstraction level to alleviate the burden of programmers and to ease reasoning about program semantics.

The rest of this paper is organized as follows. Section 2 provides some preliminaries on two major knowledge-mining tasks, i.e. classification and association mining. Section 3 proposes the medical expert system design framework with the knowledge-mining component. Running examples on medical data set and the illustration on knowledge deployment are presented in Section 4. Section 5 discusses related work and then conclusions are drawn in Section 6. The implementation of knowledge-mining component is presented in the Appendix.

## 2. Preliminaries on Tree-based Classification and Association Mining

Decision tree induction [21] is a popular method for mining knowledge from medical data and representing the result as a classifier tree. Popularity is due to the fact that mining result in a form of decision tree is interpretability, which is more concern among medical practitioners than a sophisticated method but lack of understandability. A decision tree is a hierarchical structure with each node contains decision attribute and node branches corresponding to different attribute values of the decision node. The goal of building decision tree is to partition data with mixing classes down the tree until each leaf node contains data with pure class.

In order to build a decision tree, we need to choose the best attribute that contributes the most towards partitioning data to the purity groups. The metric to measure attribute's ability to partition data into pure class is *Info*, which is the number of bits required to encode a data mixture. The metric *Info* of positive (p) and negative (n) data mixture can be calculates as:

$$Info(P(p), P(n)) = -P(p)log_2 P(p) - P(n)log_2 P(n).$$

The symbols $P(p)$ and $P(n)$ are probabilities of positive and negative data instances, respectively. The symbol $p$ represents number of positive data instances, and $n$ is the negative cases. To choose the best attribute we have to calculate information gain, which is the yield we obtained from choosing that attribute. The information gain calculation of data with two classes (positive and negative) is given as:

$$Gain(Attribute) = Info\{p/(p+n), n/(p+n)\} - \Sigma_{i=1\ to\ v}\{(p_i+n_i)/(p+n)\ Info\{\ p_i\ /(\ p_i+n_i),\ n_i\ /(\ p_i+n_i)\ \}.$$

The information gain calculates yield on *Info* of data set before splitting and *Info* after choosing attribute with $v$ splits. The gain value of each candidate attribute is calculated, and then the maximum one has been chosen to be the decision node. The process of data partitioning continues until the data subset has the same class label.

Classification task based on decision-tree induction predicts the value of a target attribute or class, whereas association-mining task is a generalization of classification in that any attribute in the data set can be a target attribute. Association mining is the discovery of frequently occurred relationships or correlations between attributes (or items) in a database. Association mining problem can be decomposed as (1) find all sets of items that are frequent patterns, (2) use the frequent patterns to generate rules. Let $I = \{i_1, i_2, i_3, ... , i_m\}$ be a set of $m$ items and $DB = \{ C_1, C_2, C_3, ..., C_n\}$ be a database of $n$ cases and each case contains items in *I*.

A *pattern* is a set of items that occur in a case. The number of items in a pattern is called the length of the pattern. To search for all valid patterns of length 1 up to $m$ in large database is computational expensive. For a set $I$ of $m$ different items, the search space of all distinct patterns can be as huge as $2^m-1$. To reduce the size of the search space, the *support* measurement has been introduced [1]. The function *support*(P) of a pattern $P$ is defined as a number of cases in DB containing *P*. Thus,

$$support(P) = |\{T \mid T \in DB,\ P \subseteq T\ \}|.$$

A pattern $P$ is called *frequent pattern* if the support value of $P$ is not less than a predefined minimum support threshold *minS*. It is the *minS* constraint that helps reducing the computational complexity of frequent pattern generation. The *minS* metric has an anti-monotone property such that if the pattern contains an item that is not frequent, then none of the pattern's supersets are frequent. This property helps reducing the search space of mining frequent patterns in algorithm Apriori [1]. In this paper we adopt this algorithm as a basis for our implementation of association mining engine.

## 3. Medical Expert System Framework and the Knowledge Mining Engines

3.1 System Architecture

Health information is normally distributive and heterogeneous. Hence, we design the medical expert system (Figure 1) to include data integration component at the top level to collect data from distributed databases and also from documents in text format.

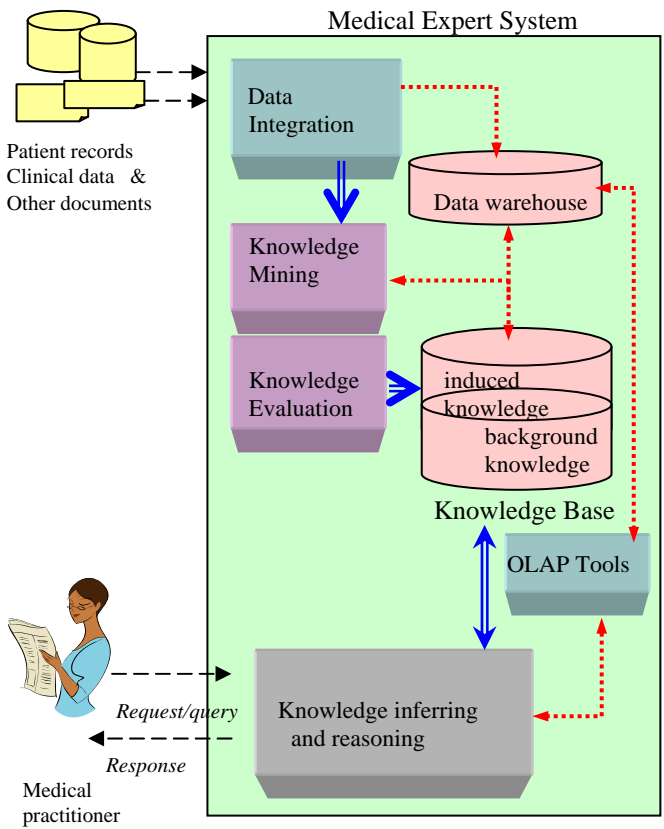

Fig. 1 Knowledge-mining component and a medical expert system framework. Double line arrows are process flow, whereas the dash line arrows are data flow.

The data integration component has been designed to input and select data with natural language processing. Data at this stage are to be stored in a warehouse to support direct querying (through OLAP tools) as well as to perform analyzing with knowledge mining engine.

Knowledge base in our design stores both induced knowledge in which its significance has to be evaluated by the domain expert, and background knowledge encoded from human experts. Knowledge inferring and reasoning is the module interfacing with medical practitioners and physicians at the front-end and accessing knowledge base at the back-end. The focus of this paper is on the implementation of knowledge-mining component, which currently contains classification and association mining engine.

3.2 Classification Mining Tool

Our classification mining engine is the implementation of decision-tree induction (ID3) algorithm [21]. The steps in our implementation are presented as follows:

**Algorithm 1** Classification mining engine
  **Input:** a data set formatted as Prolog clauses
  **Output:** a decision tree with node and edge structures
(1) Initialization
    (1.1) Clear temporary knowledge base (KB) by removing all information regarding the predicates node, edge and current_node
    (1.2) Set node counter = 0
    (1.3) Scan data set to get information about data attributes, positive instances, negative instances, total data instances
(2) Building tree
    (2.1) Increment node counter
    (2.2) Repeat steps 2.2.1-2.2.4 until there is no more attributes left for creating decision attributes
        (2.2.1) Compute the Info value of each candidate attribute
        (2.2.2) Choose the attribute that yields minimum Info to be decision node
        (2.2.3) Assert edge and node information into the knowledge base
        (2.2.4) Split data instances along node branches
    (2.3) Repeat steps 2.1 and 2.2 until the lists of positive and negative instances are empty
    (2.4) Output a tree structure that contains node and edge predicates

The program source code is based on the syntax of SWI prolog (www.swi-prolog.org).

```
main :-
   init(AllAttr, EdgeList), % initialize node
                           % and edge structures
   getNode(N),             % get node sequence number
   create_edge(N, AllAttr, EdgeList),
                           % recursively create tree
   print_model.            % print tree model
```

Classification mining engine is composed of two files *main* and *id3*. The main module (main.pl) calls initialization procedure (init) and starts creating edges and nodes of the decision tree. The data (data.pl) to be used by main module to create decision tree is also in a format of Prolog file. The mining engine induces data model of two classes: positive (class = yes) and negative (class = no). Binary classification is a typical task in medical domain. The code can be easily modified to classify data with more than two classes.

3.3 Association Mining Tool

The implementation of association mining engine is based primarily on the concept of higher-order Horn clauses. Such concept has been utilized through the predicates *maplist*, *include*, and *setof*.

The extensive use of these predicates contributes significantly to program conciseness and the ease of program verification. The program produces frequent patterns as a set of co-occurring items. To generate a nice representation of association rule such as *X => Y*, the list *L* in the predicate *association_mining* has to be further processed.

```
association_mining :-
    min_support(V), % set minimum support
    makeC1(C),     % create candidate 1-itemset
    makeL(C,L),    % compute large itemset
    apriori_loop(L,1). % recursively run apriori

makeC1(Ans):-
    input(D),  % input data as a list
    allComb(1, ItemList, Ans2),
            % make combination of itemset
    maplist(countSS(D),Ans2,Ans).
            % scan database and pass countSS
            % to maplist

makeC(N, ItemSet, Ans) :-
    input(D),  allComb(2, ItemSet, Ans1),
    maplist(flatten, Ans1, Ans2),
    maplist(list_to_ord_set, Ans2, Ans3),
    list_to_set(Ans3, Ans4),
    include(len(N), Ans4, Ans5),  % include is
            % also a higher-order predicate
    maplist(countSS(D), Ans5, Ans).
            % scan database to find: List+N
```

## 4. Running Examples and Knowledge Deployment

To show the running examples of our program coding, we use the following simple medical data represented as a Prolog file.

```
%% Data set: Allergy diagnosis
   % Symptoms of disease and their possible values
attribute( soreThroat, [yes, no]).
attribute( fever, [yes, no]).
attribute( swollenGlands, [yes, no]).
attribute( congestion, [yes, no]).
attribute( headache, [yes, no]).
attribute( class, [yes, no]).
   % Data instances
instance(1, class=no,  [soreThroat=yes, fever=yes,
          swollenGlands=yes, congestion=yes,
          headache=yes]).
instance(2, class=yes, [soreThroat=no,  fever=no,
          swollenGlands=no,  congestion=yes,
          headache=yes]).
instance(3, class=no,  [soreThroat=yes, fever=yes,
          swollenGlands=no,  congestion=yes,
          headache=no]).
…
```

Data as shown are patient records suffering from allergy (class=yes). There are ten patient records in this simple data set: patient IDs 2, 6, and 8 are those who are suffering from allergy, whereas patient IDs 1, 3, 4, 5, 7, 9, 10 are suffering from other diseases but has shown some basic symptoms similar to allergy patients. To induce classification model for allergy patients from this data, we have to save this data set as a Prolog file (data.pl) and include this file name at the header declaration of the main program. By calling predicate *main*, the system should respond as *true*. At this moment we can view the tree model by calling *listing(node)*, then *listing(edge)* and get the following results.

```
1 ?- main.
true.
2 ?- listing(node).
:- dynamic user:node/2.
user:node(1, [2, 6, 8]-[1, 3, 4, 5, 7, 9, 10]).
user:node(2, []-[1, 3, 5, 9, 10]).
user:node(3, [2, 6, 8]-[4, 7]).
user:node(4, []-[4, 7]).
user:node(5, [2, 6, 8]-[]).
true.
3 ?- listing(edge).
:- dynamic user:edge/3.
user:edge(0, root-nil, 1).
user:edge(1, fever-yes, 2).
user:edge(1, fever-no, 3).
user:edge(3, swollenGlands-yes, 4).
user:edge(3, swollenGlands-no, 5).
true.
```

The node and edge structures have the following formats:
   *node(nodeID, [Positive_Cases]-[Negative_Cases])*
   *edge(ParentNode, EdgeLabel, ChildNode)*

The node structure is a tuple of *nodeID* and a mixture of positive and negative cases represented as a list pattern: *[Positive_Cases]-[Negative_Cases]*. Node 0 is a special node, representing root node of the tree. Node 1 contains a mixture of ten patients, whereas node 5 is a pure group of allergy patients. The edges leading from node 1 to node 5 capture the model of allergy patients. Therefore, the classification result represents the following data model:
   *class(allergy) :- fever=no, swollenGlands=no.*

This model is represented as a Horn clause, thus, it provide flexibility of including this clause as a rule to select data in other group of patients who are suffering from throat infection. This kind of infection shows the same basic symptoms as allergy; therefore, screening data with the above rule can help focusing only on throat infection cases.

Applying the same data set with association mining and setting minimum support value = 50%, we got the following frequent patterns:

{fever=yes & class=no}
{fever=yes & congestion=yes}
{swollenGlands=no & congestion=yes}
{congestion=yes & headache=yes}
{congestion=yes & class=no}
{fever=yes & congestion=yes & class=no}

The first pattern can be interpreted as association rule as "*if patient has fever, that the patient does not suffer from allergy.*" This kind of rule can help accurately diagnosing patients with symptoms very close to allergy.

*Knowledge Deployment: Example 1.*
We suggest that such discovered rules, after confirming their correctness by human experts, can be added into the database system as trigger rules (Figure 2). The triggers guard database content against any updates that violates the rules. Any attempt to insert violating data will raise an error message to draw attention from the database administrator. Such trigger rules are thus deployed as a tool to enforce database integrity checking.

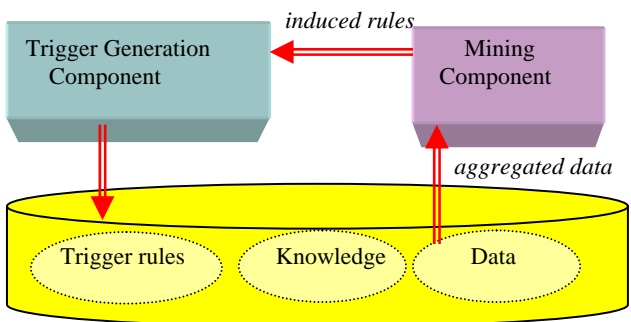

Fig. 2 The framework of knowledge deployment as triggers in a medical database.

Fig. 3 The content of automatically induced knowledge base.

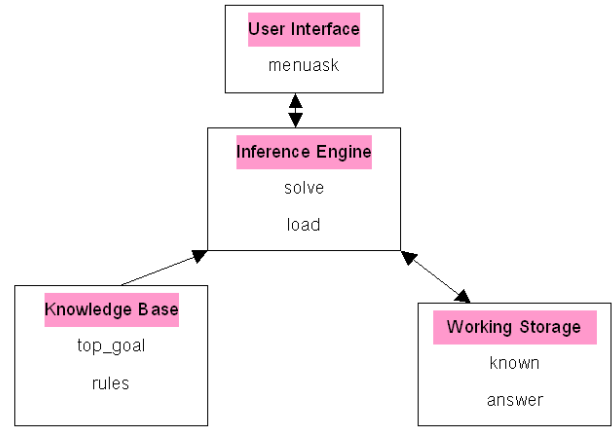

Fig. 4 Structure of a simple expert system shell with the induced knowledge base.

Fig. 5 A snapshot of medical expert system inductively created from the allergy data set.

*Knowledge Deployment: Example 2.*
The induced knowledge once confirmed by the domain expert can be added to the knowledge base of the expert system shell. We illustrate the knowledge base that automatically created from the induced tree in Figure 3. This expert system shell has simple structure as diagrammatically shown in Figure 4. User can interact with the system through a line command as shown in Figure 5, in which the user can ask for further explanation by typing the 'why' command.

## 5. Related Work

In recent years we have witnessed increasing number of applications devising database technology and machine learning techniques to mine knowledge from biomedicine, clinical and health data. Roddick et al [22] discussed the two categories of mining techniques applied over medical data: explanatory and exploratory. Explanatory mining refers to techniques that are used for the purpose of confirmation or making decisions. Exploratory mining is data investigation normally done at an early stage of data analysis in which an exact mining objective has not yet been set.

Explanatory mining in medical data has been extensively studied in the past decade employing various learning techniques. Bojarczuk et al [4] applied genetic programming method with constrained syntax to discover classification rules from medical data sets. Thongkam et al [28] studied breast cancer survivability using AdaBoost algorithm. Ghazavi and Liao [9] proposed the idea of fuzzy modeling on selected features of medical data. Huang et al [11] introduced a system to apply mining techniques to discover rules from health examination data. Then they employed a case-based reasoning to support the chronic disease diagnosis and treatments. The recent work of Zhuang et al [31] also combined mining with case-based reasoning, but applied a different mining method. They performed data clustering based on self-organizing maps in order to facilitate decision support on solving new cases of pathology test ordering problem. Biomedical discovery support systems are recently proposed by a number of researchers [5, 6, 10, 29, 30]. Some work [20, 25] extended medical databases to the level of data warehouses.

Exploratory, as oppose to explanatory, is rarely applied to medical domains. Among the rare cases, Nguyen et al [19] introduced knowledge visualization in the study of hepatitis patients. Palaniappan and Ling [20] applied the functionality of OLAP tools to improve visualization in data analysis.

It can be seen from the literature that most medical knowledge discovery systems have applied only some mining techniques to discover hidden knowledge with the main purpose to support medical diagnosis [4, 14, 17]. Some researchers [3, 8, 15, 16] have extended the knowledge discovery aspect to the large scale of a medical decision support system.

Our work is also in the main stream of medical decision support system development, but our methodology is different from those appeared in the literature. The system proposed in this paper is based on a logic-programming paradigm. The justification of our logic-based system is that the closed form of Horn clauses that treats program in the same way as data facilitates fusion of knowledge learned from different sources, which is a normal setting in medical domain. Knowledge reuse can easily practice in this framework.

The declarative style of our implementation also eases the future extension of the proposed medical support system to cover the concepts of higher-order mining [23], i.e. mining from the discovered knowledge, and constraint mining [7], i.e. mining with some specified constraints to obtain relevant knowledge.

## 6. Conclusions and Discussion

Modern healthcare organizations generate huge amount of electronic data stored in heterogeneous databases. Data collected by hospitals and clinics are not yet turned into useful knowledge due to the lack of efficient analysis tools. We thus propose a rapid prototyping of automatic mining tools to induce knowledge from medical data. The induced knowledge is to be evaluated and integrated into the knowledge base of a medical expert system. Discovered knowledge facilitates the reuse of knowledge base among decision-support applications within organizations that own heterogeneous clinical and health databases. Direct application of the proposed system is for medical related decision-making. Other indirect but obvious application of such knowledge is to pre-process other data sets by grouping it into focused subset containing only relevant data instances.

The main contribution of this work is our implementation of knowledge mining engines based on the concept of higher-order Horn clauses using Prolog language. Higher-order programming has been originally appeared in functional languages in which functions can be passed as arguments to other functions and can also be returned from other functions. This style of programming has soon been ubiquitous in several modern programming languages such as Perl, PHP, and JavaScript. Higher order style of programming has shown the outstanding benefits of code reuse and high level of abstraction. This paper illustrates higher order programming techniques in SWI-Prolog. The powerful feature of meta-level programming in Prolog facilitates the reuse of mining results represented as rules to be flexibly applied as conditional clauses in other applications.

The plausible extensions of our current work are to add constraints into the knowledge mining method in order to limit the search space and therefore yield the most relevant and timely knowledge, and due to the uniform representation of Prolog's statements as a clausal form, mining from the previously mined knowledge should be implemented naturally. We also plan to extend our system to work with stream data that normally occur in modern medical organizations.

**Appendix**

The implementation of knowledge-mining component is based on the concept of higher-order and meta-programming styles. Higher-order programming in Prolog refers to Horn clauses that can quantify over other predicate symbols [18]. Meta-level programming is also another powerful feature of Prolog. Data and program in Prolog take the same representational format; that is clausal form. Higher-order and meta-level clauses in the following source code are typed in bold face.

```prolog
/* Classification mining engine */
:- include('data.pl').
:- dynamic current_node/1, node/2, edge/3.

main :-
    init(AllAttr, EdgeList),
    getNode(N),           % get node sequence number
    create_edge(N, AllAttr, EdgeList),
    print_model.

init(AllAttr, [root-nil/PB-NB]) :-
    retractall(node(_, _)),
    retractall(current_node(_)),
    retractall(edge(_, _, _)),
    assert(current_node(0)),
    findall(X, attribute(X, _), AllAttr1),
    delete(AllAttr1, class, AllAttr),
    findall(X2, instance(X2, class=yes, _), PB),
    findall(X3, instance(X3, class=no, _), NB).

getNode(X) :-
    current_node(X), X1 is X+1,
    retractall(current_node(_)),
    assert(current_node(X1)).

create_edge(_, _, []) :- !.

create_edge(_, [], _) :- !.

create_edge(N, AllAttr, EdgeList) :-
    create_nodes(N, AllAttr, EdgeList).

create_nodes(_, _, []) :- !.

create_nodes(_, [], _) :- !.

create_nodes(N, AllAttr, [H1-H2/PB-NB|T]) :-
    getNode(N1),   % get node sequence number N1
    assert(edge(N, H1-H2, N1)), % H1-H2 is
                                % a pattern
    assert(node(N1, PB-NB)),    % PB-NB is
                                % a pattern
    append(PB, NB, AllInst),
    ((PB \== [], NB \== []) ->  % if-condition
                                % then clauses
        (cand_node(AllAttr, AllInst, AllSplit),
         best_attribute(AllSplit, [V, MinAttr,
                                         Split]),
         delete(AllAttr, MinAttr, Attr2),
         create_edge( N1, Attr2, Split))
        ;                       % else clause
         true ),
    create_nodes(N, AllAttr, T).
%

% select best attribute to be a decision node
%
best_attribute([], Min, Min).

best_attribute([H|T], Min) :-
    best_attribute(T, H, Min).

best_attribute([H|T], Min0, Min) :-
    H = [V, _, _ ],
    Min0 = [V0, _, _ ],
    ( V < V0 -> Min1 = H ; Min1 = Min0),
    best_attribute(T, Min1, Min).

%
% generate candidate decision node
%
cand_node([], _, []) :- !.

cand_node(_, [], []).

cand_node([H|T], CurInstL, [[Val,H,SplitL]
                            |OtherAttr]) :-
    info(H, CurInstL, Val, SplitL),
    cand_node(T, CurInstL, OtherAttr).

%
% compute Info of each candidate node
%
info(A, CurInstL, R, Split) :-
    attribute(A,L),
    maplist(concat3(A,=), L, L1),
    suminfo(L1, CurInstL, R, Split).

concat3(A,B,C,R) :-
    atom_concat(A,B,R1),
    atom_concat(R1,C,R).

suminfo([],_,0,[]).

suminfo([H|T], CurInstL, R, [Split | ST]) :-
    AllBag = CurInstL, term_to_atom(H1, H),
    findall(X1, (instance(X1, _, L1),
                 member(X1, CurInstL),
                 member(H1, L1)), BagGro),
    findall(X2,(instance(X2, class=yes, L2),
                 member(X2, CurInstL),
                 member(H1, L2)), BagPos),
    findall(X3,(instance(X3, class=no, L3),
                 member(X3, CurInstL),
                 member(H1, L3)), BagNeg),
    (H11= H22) = H1,
    length(AllBag, Nall),
    length(BagGro, NGro),
    length(BagPos, NPos),
    length(BagNeg, NNeg),
    Split = H11-H22/BagPos-BagNeg,
    suminfo(T, CurInstL, R1,ST),
    ( NPos is 0 *-> L1 = 0;
                L1 is (log(NPos/NGro)/log(2)) ),
    ( 0 is NNeg *-> L2 = 0;
                L2 is (log(NNeg/NGro)/log(2)) ),
    ( NGro is 0 -> R= 999;
                R is (NGro/Nall)*
                        (-(NPos/NGro)*
                        L1- (NNeg/NGro)*L2)+R1).

/* ======================= */
/* Association mining engine */
```

```
/* ========================= */
association_mining:-
    min_support(V),   % set minimum support
    makeC1(C), % create candidate 1-itemset
    makeL(C,L),% compute large itemset
    apriori_loop(L,1). % recursively run apriori

apriori_loop(L, N) :- % base case of recursion
    length(L) is 1,!.

apriori_loop(L, N) :- % inductive step
    N1 is N+1,
    makeC(N1, L, C),
    makeL(C, Res),
    apriori_loop(Res, N1).

makeC1(Ans):-
    input(D), % input data as a list,
              %  e.g. [[a], [a,b]]
              % then make combination of itemset
    allComb(1, ItemList, Ans2),
       % scan database and pass countSS to maplist
    maplist(countSS(D),Ans2,Ans).

makeC(N, ItemSet, Ans) :- input(D),
    allComb(2, ItemSet, Ans1),
    maplist(flatten, Ans1, Ans2),
    maplist(list_to_ord_set, Ans2, Ans3),
    list_to_set(Ans3, Ans4),
    include(len(N), Ans4, Ans5),
                % include is also a
                % higher-order predicate
    maplist(countSS(D), Ans5, Ans).
         % scan database to find: List+N

makeL(C, Res):-    % for all large itemset creation
                   % call higher-order predicates
                   % include and maplist
    include(filter, C, Ans),
    maplist(head, Ans, Res).

%
% filter and head are for pattern matching of
%   data format
%
filter(_+N):-
    input(D),
    length(D,I),
    min_support(V),
    N>=(V/100)*I.

head(H+_, H).

%
% an arbitrary subset of the set containing
% given number of elements
%
comb(0, _, []).

comb(N, [X|T], [X|Comb]) :-
    N>0, N1 is N-1,
    comb(N1, T, Comb).

comb(N, [_|T], Comb) :-
    N>0,
    comb(N, T, Comb).
allComb(N, I, Ans) :-
    setof(L, comb(N, I, L), Ans).

countSubset(A, [], 0).

countSubset(A, [B|X], N) :-
    not(subset(A, B)),
    countSubset(A, X, N).

countSubset(A, [B|X], N) :-
    subset(A, B),
    countSubset(A, X, N1),
    N is N1+1.

countSS(SL, S, S+N) :-
    countSubset(S, SL, N).
```

## Acknowledgments

This work has been fully supported by research fund from Suranaree University of Technology granted to the Data Engineering and Knowledge Discovery (DEKD) research unit. This research is also supported by grants from the National Research Council of Thailand (NRCT) and the Thailand Research Fund (TRF).

**Nittaya Kerdprasop** is an associate professor at the school of computer engineering, Suranaree University of Technology, Thailand. She received her B.S. in radiation techniques from Mahidol University, Thailand, in 1985, M.S. in computer science from the Prince of Songkla University, Thailand, in 1991 and Ph.D. in computer science from Nova Southeastern University, USA, in 1999. She is a member of IAENG, ACM, and IEEE Computer Society. Her research of interest includes Knowledge Discovery in Databases, Data Mining, Artificial Intelligence, Logic and Constraint Programming, Deductive and Active Databases.

**Kittisak Kerdprasop** is an associate professor and the director of DEKD (Data Engineering and Knowledge Discovery) research unit at the school of computer engineering, Suranaree University of Technology, Thailand. He received his bachelor degree in Mathematics from Srinakarinwirot University, Thailand, in 1986, master degree in computer science from the Prince of Songkla University, Thailand, in 1991 and doctoral degree in computer science from Nova Southeastern University, USA, in 1999. His current research includes Data mining, Machine Learning, Artificial Intelligence, Logic and Functional Programming, Probabilistic Databases and Knowledge Bases.